# Two-Dimensional Electron Gas as a Basis for Low-Loss Hyperbolic Metamaterials

Michael A. Mastro
US Naval Research Lab

**Abstract**
The implementation of hyperbolic metamaterials as component in optical waveguides, semiconductor light emitters and solar cells has been limited by the inherent loss in the metallic layers. The features of a hyperbolic metamaterial arise by the presence of alternating metal and a dielectric layers. This work proposes that the deleterious loss characteristic of metal-based hyperbolic metamaterials can be minimized by employing a III-nitride superlattice wherein a two-dimensional electron gas (2DEG) functions as the metallic layer.

**Introduction**
Evolving developments in hyperbolic metamaterials are now poised to make major advancements in critical components of optoelectronics and active nano-optics. The typical design of a hyperbolic metamaterial consists of alternating layers of a metal and a dielectric. This structure possesses an asymmetry in its perpendicular and parallel permittivity components that manifests as a hyperboloid in the dispersion curve.

Dramatic differences in the electromagnetic response can be seen depending on the free-space wavelength, material parameters such as the metal/dielectric ratio, and the electron density of the metallic and the dielectric layers. Specifically, hyperbolic metamaterials present type-1 and type-2 metamaterial regions as well as metal-like and dielectric-like regions.

We propose to utilize a two-dimensional electron gas (2DEG) formed at a semiconductor interface as the metallic layer and the low-doped semiconductor layers as the dielectric in the hyperbolic metamaterial. A 2DEG is formed in materials such as AlGaAs/GaAs by bandgap engineering wherein doping in the AlGaAs barrier creates carriers that spill into a potential well (at the AlGaAs/GaAs interface) that is spatially displaced from the ionized donors. This displacement allows the carriers to experience few scattering events and thus move with a high mobility.

A related design in the III-nitride system relies on the large polarization fields that are present at the heterostructure interfaces. This article proposes the use of a III-nitride superlattice (e.g., AlN/GaN) as a low-loss pseudo-metal / dielectric hyperbolic metamaterial.

**Theoretical**
Hyperbolic metamaterials possess components of opposite sign in the permittivity tensor,

$$\varepsilon = \begin{bmatrix} \varepsilon_{xx} & 0 & 0 \\ 0 & \varepsilon_{yy} & 0 \\ 0 & 0 & \varepsilon_{zz} \end{bmatrix},$$

where the in-plane components themselves are equal $\varepsilon_{xx} = \varepsilon_{yy}$ but unequal to the perpendicular component $\varepsilon_{zz}$, that is, $\varepsilon_{zz} \neq \varepsilon_{xx} = \varepsilon_{yy}$. [1]
The opposing sign in the components in the permittivity tensor, $\varepsilon_{zz}\varepsilon_{xx} < 0$, creates a hyperbolic isofrequency curve in k-space as described by

$$\frac{k_x^2 + k_y^2}{\varepsilon_{zz}} + \frac{k_z^2}{\varepsilon_{xx} + \varepsilon_{yy}} = \frac{\omega^2}{c^2} = k_0^2,$$

where $k_0$ is the free space wavevector, $\omega$ is the angular frequency, and c is the speed of light. For example, figure 1a depicts a type-2 hyperbolic metamaterial where $\varepsilon_{xx} = \varepsilon_{yy} < 0$ and $\varepsilon_{zz} > 0$. In a type-2 hyperbolic metamaterial, electromagnetic waves can propagate in-plane with very large wavevectors. Ignoring the $k_y$ plane for illustration and setting the $k_z$ component to zero, results in the minimum allowed $k_x$ wavevector (for a fixed frequency) to be

$$k_{x-\min}^2 = \varepsilon_{zz}\frac{\omega^2}{c^2} = \varepsilon_{zz} k_0^2.$$

This hyperbolic isofrequency curve is in contrast to the circular isofrequency curve present in most natural materials This has interesting consequence when one considers the same information in the $k_x : k_z$ plane in figure 1a. The circular isofrequency curves of air (free space) and in GaN are also presented.

Light propagating in air with a certain angle relative to material slab (and thus $k_x$ wavevector) can only propagate into another material by matching the transverse wavevectors, e.g., $k_{x-Air} = k_{x-GaN}$. Light traveling from air at any angle relative to the GaN slab can propagate into the GaN given that a transverse component in GaN, $k_{x-GaN}$, on the isofrequency curve is accessible for all possible transverse components in air, $k_{x-Air}$. In contrast, only light propagating in GaN within a narrow angular cone relative to the air/GaN interface can propagate into the air by matching an available transverse wavevector in air. Light propagating in GaN at a larger angle relative to the surface normal cannot match the transverse wavevector and thus cannot propagate into the air, and will suffer from total internal reflection.

Examining figure 1a, it is evident that any electromagnetic wave propagating in the type-2 hyperbolic metamaterial cannot match the transverse wavevector of air (or GaN) and cannot propagate into the air (or GaN). This is advantageous for constructing optical waveguides for in-plane propagation without loss into the surrounding air or material (e.g., GaN, $SiO_2$, Si). This could occur when launching an electromagnetic wave into a waveguide is possible by building the emitter inside the metamaterial, employing a grating, or by relying on evanescent coupling. The small mode volume of large wavevectors enables tight confinement in metamaterial waveguides.

For a metamaterial laser, the small mode volume of large wavevectors would allow this metamaterial to act as an optical cavity where the optical mode can be sub-diffraction in size. The active semiconductor region can be located adjacent to or within the metamaterial. This design can allow amplified spontaneous emission or thresholdless lasing.

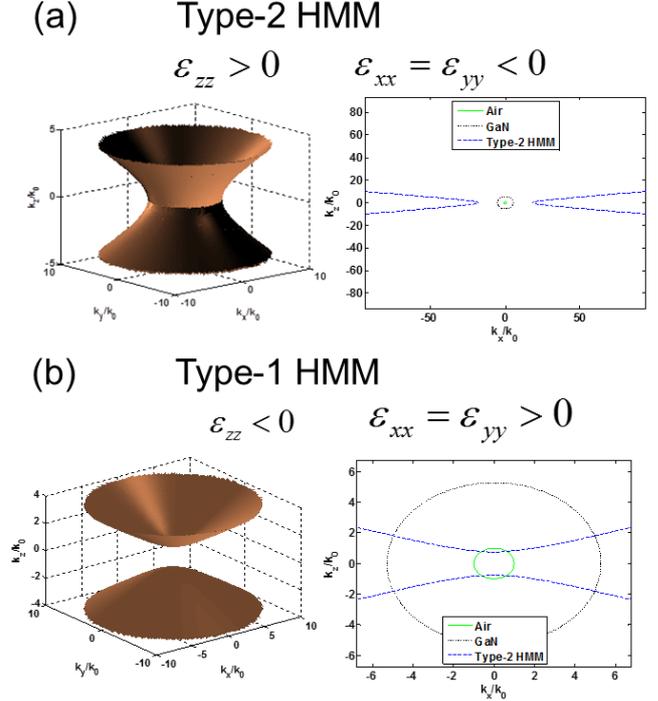

Figure 1. Isofrequency contour of a (a) type-2 and (b) type-1 hyperbolic metamaterial contrasted to homogeneous GaN in (left) three-dimensional k-space and (right) a corresponding two-dimensional slice where $k_y = 0$. In a typical homogeneous material, the permittivity, ε, is the same in all directions, i.e., a sphere in k-space (for constant frequency). In contrast, a hyperbolic metamaterial presents a permittivity and thus light propagation different in the perpendicular (z) and parallel (x or y) directions – relative to the surface. When only the parallel or perpendicular permittivity is negative (e.g., $\varepsilon_{zz}<0$ and $\varepsilon_{xx,yy}>0$; or $\varepsilon_{zz}>0$ and $\varepsilon_{xx,yy}<0$) then the k-space isofrequency curve is a hyperboloid.

Another use of this forbidden propagation is to use the type-2 hyperbolic metamaterial as a reflector in a light emitting structure. This simply could be used to reflect light generated in the active region away from a lossy substrate such as Si. [2] A more sophisticated technique would be to use the type-2 hyperbolic metamaterial as a reflective surface as part of the cavity that surrounds the light emitting layers as in a resonant cavity light emitting diode [3] or a polariton laser. [4]

Figure 1b shows the structure of a type-1 hyperbolic metamaterial where $\varepsilon_{xx} = \varepsilon_{yy} > 0$ and $\varepsilon_{zz} < 0$. In a type-1 hyperbolic metamaterial,

$$k_z^2 = \varepsilon_{xx}\left[k_0^2 - \frac{k_x^2 + k_y^2}{\varepsilon_{zz}}\right],$$

is parabolic. Large transverse wavevectors are present in the type-1 hyperbolic metamaterial. It should be noted that the parabolic relation does not hold as the wavevector approaches the physical dimensions of the unit cell with resultant diffraction effects. A type-1 hyperbolic metamaterial has a high density of states, which can prove advantageous for enhancement of the radiative rate of an emitter via a broadband Purcell effect. This enhancement in the radiative rate is advantageous for emitters used in communication which require a rapid cycling of on/off states. Power flow is described by the Poynting vector, which is perpendicular to the dispersion curve as shown in the dispersion curve in figure 1. Therefore, the power flow is contained within the asymptotes of the perpendicular hyperbola as set by the angular cone given by $\tan(\theta) = \sqrt{\varepsilon_{xx}/|\varepsilon_{zz}|}$. A type-1 metamaterial can used for highly directional emission where the beam is confined to a subdiffraction cone when $|\varepsilon_{zz}| >> \varepsilon_{xx}$. [5]

Hyperbolic metamaterials are formed as a periodic metal / dielectric structure with dimensions, λ/10 to λ/100, much smaller than the wavelength of light. The two most common structures are vertical metallic nanowires in a dielectric matrix, and alternating metal / dielectric layers as a superlattice. In the planar superlattice the effective perpendicular permittivity can be expressed by

$$\varepsilon_\perp = \frac{\varepsilon_{Metal}\varepsilon_{Dielectric}}{f\varepsilon_{Metal} + (1-f)\varepsilon_{Dielectric}} = \frac{1}{\frac{d_{Metal}/\varepsilon_{Metal} + d_{Dielectric}/\varepsilon_{Dielectric}}{d_{Metal} + d_{Dielectric}}},$$

where f is the fill fraction of the metal. This equation shows that the negative effective perpendicular permittivity is possible when the permittivity of the metal is also negative, which occurs for frequencies below the plasma frequency. Metals such as silver and gold, have plasma frequencies in the visible, are relatively low-loss compared to other metals, and are thus commonly used as the metallic component in the metamaterial.

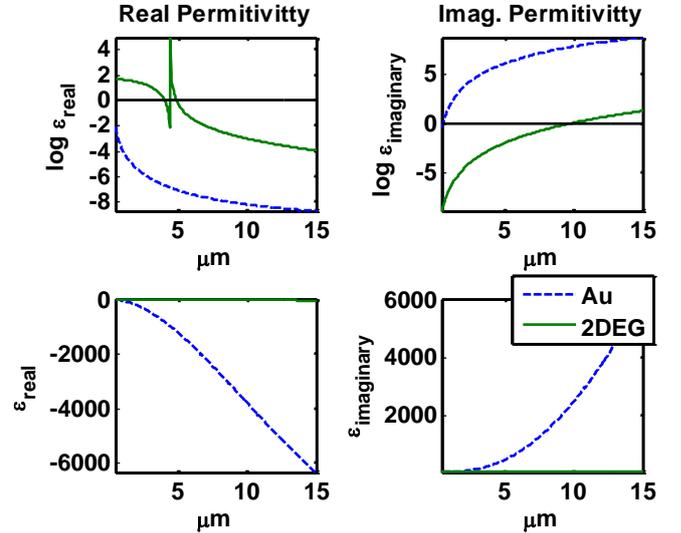

Figure 2. The (left) real and (right) imaginary permittivity response of a 4x10[13] carrier concentration 2DEG layer in GaN and the permittivity of gold. Note that the log plots (top) were multiplied with a signed component to match the negative values in the linear plots (bottom).

Additionally, in the superlattice the parallel permittivity can be expressed by

$$\varepsilon_\parallel = f\varepsilon_{Metal} + (1-f)\varepsilon_{Dielectric} = \frac{d_{Metal}\varepsilon_{Metal} + d_{Dielectric}\varepsilon_{Dielectric}}{d_{Metal} + d_{Dielectric}}.$$

Here the negative permittivity of the metal must overcome the positive dielectric permittivity to achieve an effective negative parallel permittivity. According to the model of a Drude metal, the metallic permittivity becomes more negative for lower frequencies (longer wavelengths). Therefore, the parallel permittivity typically becomes negative only at wavelengths slightly longer than the metal plasma frequency.

In general, the optimal hyperbolic metamaterial would have a metallic material that exhibits little loss. Additionally, it is often beneficial if the absolute values of metal and dielectric permittivity are within one order of magnitude, so that the response of the metamaterial can be designed with metallic layers of reasonable thickness. Far below the plasma frequency, metals display a negative permittivity two or three orders of magnitude larger than the dielectric permittivity. This necessitates thin metallic layers, which are difficult to produce and often exhibit losses arising from surface roughening.

Moreover, bulk metals, even gold and silver, show fairly high loss (as represented by the imaginary component of the permittivity) near the plasma frequency and far below the plasma frequency. [6]

The Lorentz model for a metal or dielectric is

$$\varepsilon(\omega) = \varepsilon_0 + \frac{\varepsilon_0 \frac{Ne^2}{m\varepsilon_0}}{\omega_0^2 + \omega + j\gamma\omega}$$

where γ is the collisions per unit time, τ=1/γ, is the mean time between collisions, and the plasma frequency

$$\omega_p^2 = \frac{Ne^2}{m\varepsilon_0},$$

is proportional to the carrier concentration N.

Decreasing the metal plasma frequency requires diluting the number of carriers. This has been attempted by alloying the metal with a less conductive substance but generally this results in a large increase in γ and, consequently, loss. Rather than decreasing the plasma frequency in a metal, another technique is to degenerately dope a semiconductor to increase the plasma frequency. Deposition technologies such as MOCVD, MBE, and ALD can easily deposit very thin semiconductor layers with high doping levels. Heavily doping a semiconductor creates more carriers from the ionized donors (or acceptors). Nevertheless, it is these same ionized donors that create scattering centers that decrease the mobility and similarly increase the loss factor γ. Furthermore, the plasma frequency of a typical moderately doped semiconductor is in the mid-IR. Shifting the plasma frequency into the near-IR or even visible requires doping levels beyond the solubility limit of most semiconductors.

A 25nm $Al_{0.25}Ga_{0.75}N$ layer on a GaN film will create a polarization charge at the AlGaN/GaN interface that will accumulate carriers of approximately $1\times10^{13}$ electrons/cm$^2$ with a mobility greater than 1000 cm$^2$/(V-s). Various doping profiles can be used to modify the characteristics of the device but this high electron mobility transistor structure is commonly demonstrated without any intentional doping. A more advanced design employs a 3nm AlN layer on a GaN film. This structure will have a much larger polarization field at the AlN/GaN interface and will accumulate carriers up to $4\times10^{13}$ electrons/cm$^2$.

The permittivity response at the AlN/GaN near interfacial region with a carrier concentration of $4\times10^{13}$ electrons/cm$^2$ modeled according to the Lorentz equation is compared to the permittivity of gold in figure 2. The lack of scattering sites in the 2DEG modeled in the collision term γ results in much smaller imaginary component of the permittivity and, thus, loss in the 2DEG compared to gold in this wavelength range. Additionally, the real component of the permittivity of the 2DEG is within two orders of magnitude of a typical dielectric, which is beneficial for the fabrication of realistically sized layer in a hyperbolic metamaterial.

Figure 3 shows an effective medium calculation of this structure for a 2DEG with a sheet density of $4\times10^{13}$ electrons/cm$^2$. Specifically, figure 3-top displays the regions of positive and negative effective perpendicular (or zz notation) permittivity (figure 3-top), figure 3-middle displays the regions of positive and negative effective parallel (or xx or yy notation) permittivity, and figure 3-bottom combines this information to display the regions of metallic, dielectric, type-1 hyperbolic metamaterial, and type-2 hyperbolic metamaterial response. Figure 3 is shown as a function of the 2DEG thickness fraction in each period of the superlattice. The effective 2DEG thickness is the determined by the width of the potential well. A simple AlN/GaN interface will have a triangular potential on the GaN side of the interface where most of the carriers are contained within the first three nanometers. [7]

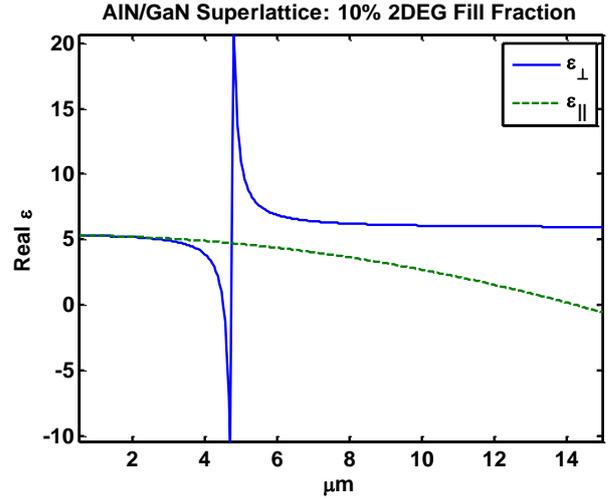

Figure 4. Effective medium analysis of an AlN/GaN superlattice with a carrier concentration in the 2DEG of $4\times10^{13}$ electrons/cm$^2$ for a 10% 2DEG thickness fraction.

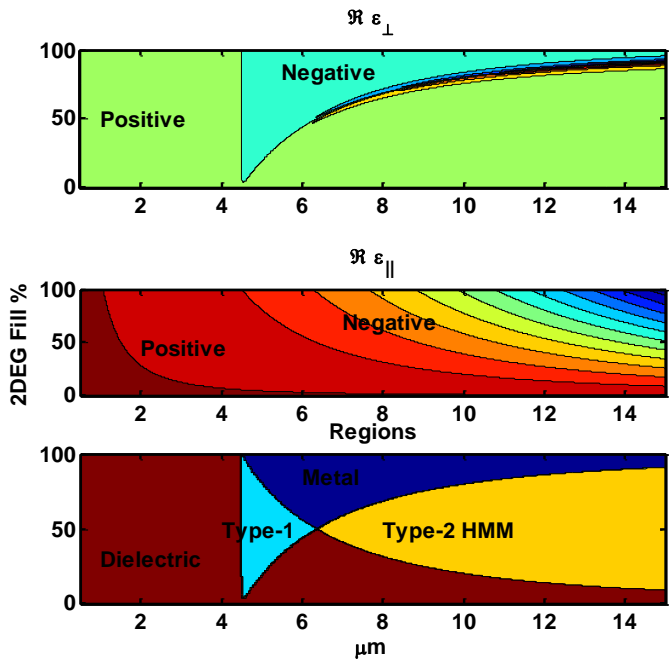

Figure 3. Effective medium analysis of an AlN/GaN superlattice with a carrier concentration in the 2DEG of $4\times10^{13}$ electrons/cm$^2$ as a function of the thickness fill fraction of the 2DEG. A type-1 hyperbolic metamaterial is present when $\varepsilon_{xx}=\varepsilon_{yy}>0$ and $\varepsilon_{zz}<0$; a type-2 hyperbolic metamaterial is present when $\varepsilon_{xx}=\varepsilon_{yy}<0$ and $\varepsilon_{zz}>0$.

Figure 4 displays a two-dimensional slice of figure 3-bottom for a fill fraction of 10%. A type-1 hyperbolic metamaterial is evident at 4.5µm and a type-2 hyperbolic metamaterial is evident at 15µm.

## Discussion

The possibility to deposit a 2DEG superlattice in-situ opens a number of interesting designs. For example, depositing the 2DEG-based hyperbolic metamaterial adjacent to a quantum well emitter is a promising design for strong-coupling. Strong-coupling is only achievable if the non-radiative processes are minimal and injection is efficient.

Normally, strong-coupling involves coupling the emitter to the narrow-band modes of a cavity. In contrast, the modes of the hyperbolic metamaterial could efficiently couple to the emitter – without the need for a cavity. One application of efficient strong coupling is the polariton laser wherein the bound electron-hole pair indistinguishably couples to the optical state to create an exciton-polariton. The polariton decays as a coherent directional beam of photons similar to a laser but with essentially thresholdless operation. Room-temperature operation is reasonably achievable in only GaN owing to its unusually large exciton binding energy.

The type-1 and type-2 metamaterial regions differ although both display a high-density of states bound only by the experimentally achievable thinness and smoothness of the layers. Forming an emitter such as a quantum well in proximity to a hyperbolic metamaterial will dramatically increase the radiative rate and alter the emission pattern. This irreversible weak-coupling provides additional decay routes to compete against the non-radiative processes.

Recent theoretical developments have shown that Van Hove singularities can exist in a hyperbolic metamaterials. These slow-light states can be used to dramatically increase the photonic density of states. Analogous to a plasmonic laser, coupling the emitter and metamaterial would create a metamaterial laser with a hybrid sub-diffraction sized optical mode.

In general, hyperbolic metamaterials are efficient for coupling to emitters or simply acting as a passive guide of light within a sub-diffraction size spot. Employing semiconductor based 'metallic' layers such as III-nitride based 2DEGs offers the ability to modulate the guide. Gated structures can deplete or enhance the density of carriers, alter the plasma frequency, and thus change the regime of operation of the hyperbolic metamaterial to guide, absorb, or reflect the light. Particularly interesting is that this optical transition can dramatically affect the above-described weak and strong coupling to an emitter.

**Conclusion**

A design was presented to create an all semiconductor hyperbolic metamaterial based on a 2DEG formed at Al(Ga)N/GaN interfaces. This design removes a number of the inherent loss mechanisms present in traditional metal based hyperbolic metamaterials. Furthermore, this 2DEG superlattice can be gated to modulate the Fermi level, alter the number of carriers in the well, and thus electrically modulate the properties of the metamaterial.

**Acknowledgments**

Research at NRL is supported by the Office of Naval Research.